**The interplay between tissue healing and bone remodeling around immediately loaded tooth replacement implants**

Soroush Irandoust and Sinan Müftü*

Department of Mechanical and Industrial Engineering

Northeastern University

Boston, MA, USA 02115

irandoust@coe.neu.edu

Corresponding Author: s.muftu@northeastern.edu – 617.373.4743

**ABSTRACT**

Long-term bone healing/adaptation after a dental implant treatment starts with diffusion of mesenchymal stem cells to the fracture callus and their subsequent differentiation. The healing phase is followed by the bone-remodeling phase. In this work, a mechano-regulatory cellular differentiation model was used to simulate tissue healing around an immediately loaded dental implant. All tissue types were modeled as poroelastic in the healing phase. Material properties of the healing region were updated after each loading cycle for 30 cycles (days). The tissue distribution in the healed state was then used as the initial condition for the remodeling phase during which regions healed into bone adapt their internal density with respect to a homeostatic remodeling stimulus. The short- and long-term effects of micro-motion on bone healing and remodeling were studied. Development of soft tissue was observed both in the coronal region due to high fluid velocity, and on the vertical sides of the healing-callus due to high shear stress. In cases with small implant micromotion, tissue between the implant threads differentiated into bone during the healing phase, but resorbed during remodeling. In cases with large implant micromotion, higher percentage of the healing region differentiated into soft tissue resulting in less volume available for bone remodeling. But, the remaining bone region developed higher density bone tissue. It was concluded that an optimal range of controlled micromotion could be designed for a given implant in order to achieve the desired functional properties.

**INTRODUCTION**

Dental Implant fixtures have become an integral part of treatment for partially or fully edentulous patients [1] since Branemark introduced the two-stage treatment protocol [2]. A single-stage protocol, where the implant is surgically inserted, the prosthetic tooth installed and the implant immediately loaded, is considered beneficial as it reduces the number of surgical interventions. Osseointegration of immediately loaded implants has been the subject of numerous clinical and animal studies. Provided that the primary stability of the implant can be ensured [3,4], immediate loading has been shown to be a reliable treatment [1,5], without disturbing the biological osseointegration process [6] or affecting bone mineral apposition rate [7]. Nevertheless, high occlusal loading is considered as a risk factor for immediately loaded implants [5].

Dental implant surgery causes a wounded region around the implant, which initiates the tissue healing process. Intramembranous bone formation starts with blood clot formation, vascularization within the fracture callus, and proliferation and migration of mesenchymal stem cells (MSCs) from surrounding bone marrow [8]. Under favorable conditions and stable sites, MSCs differentiate into osteoblasts and woven bone forms through osteogenesis [9] followed by compaction of woven bone. After about a month [8], bone remodeling starts. Bone continuously adapts itself by adjusting its mass density to mechanical loading and functionality [10,11].

Studies of bone healing around immediately loaded implants typically use displacement-controlled micromotion to assess healing pathway [12]. On the other hand, remodeling studies only consider the mastication force as the loading input [13]. It is of course interesting to note that in a micromotion-controlled environment the tissue properties change continuously and the load carrying capacity of the tissue adjusts accordingly.

Clinical and experimental examinations create the opportunity to observe biological processes of fracture healing [8,14,15] and bone remodeling [16]; and, in-silico studies can represent how biological factors contribute to the outcome of a dental implant treatment [17]. Numerous computer simulations have been carried out to investigate effects of mechanical loading on bone fracture healing and bone remodeling [13,18-27]. In healing studies, long-term adaptation of the bone tissue is not investigated, and remodeling studies usually do not start from a realistic initial state. To the best of our knowledge, this is the first study modeling both biological processes consecutively.

**RESULTS**

Transient change of elastic modulus during healing and remodeling for different micromotion ranges is shown in Fig. 1. It is interesting to note that regardless of the micromotion range, the tissue between the implant threads develop into bone during the healing phase (days 1 – 30) but resorb during remodeling. The fate of the tissue on the vertical sides of the healing callus strongly depends on the micromotion amplitude.

Fig. 2 shows the distribution of solid and fluid stimulus in the healing gap at days 5 and 30. Regions in between implant threads experience the lowest solid stimulus (lowest shear strain) and lowest fluid velocity compared to the other regions of the callus. These regions have smaller shares of transferring mechanical load to surrounding cancellous bone. This characteristic leads these regions to a faster healing during the healing phase, but to resorption later during remodeling phase. Resorption due to insufficient remodeling stimulus is known as stress shielding [28]. During the healing phase, large loading amplitude ($z_{max}$ = 20 µm) causes soft tissue development on the vertical sides of the healing callus and in the coronal region (Fig. 1). This observation is correlated with the high solid and fluid stimuli in these regions (Fig. 2). Note that the high fluid velocity in the coronal region is due to the very low permeability of the adjacent cortical bone.

In order to get a closer look on how different types of tissue evolve during the healing and remodeling phases, two metrics are defined. Tissue volume (TV) is the ratio of the volume of a specific tissue type to the volume of the healing region. Tissue-to-implant contact (TIC) shows how much of the implant surface is in contact with a specific tissue type.

The transient changes of the soft and resorbed tissue volumes and implant contact ratios (TV and TIC) are presented in Fig. 3 during the healing and remodeling phases. For this implant, very small amount of soft fibrous tissue develops during healing. Nevertheless, larger micromotion results in more of fibrous tissue. Bone resorption during the remodeling phase results in 15%, 30% and 35% of soft-tissue-TV for $z_{max}$ values of 20, 10 and 5 µm, respectively. Similar trends are seen for the soft-tissue-TIC values; $z_{max}$ values of 5 and 10 µm result in about 50% TIC and 20 µm results in about 30% TIC.

The healing and remodeling histories of the fractions (TV, TIC) of immature and trabecular bone (0.1 < E < 3.5 GPa) are presented in Fig. 4. Two important general observations can be made from these results. First, it is seen that low amplitude motion results in more bone during the healing phase. Second, it is also seen that a non-negligible volume of the bone that develops during the healing phase resorbs during remodeling. In particular, implant motion with 5 µm range results in all callus tissue to heal into bone. As the range of motion is increased to 10 and 20 µm, the TV values are reduced to 95% and approximately 40%, respectively. The area of bone making contact with the implant (TIC) shows similar trends, where 5, 10 and 20 µm result in 100%, 95% and 65% TIC values, respectively. A clear reduction in both bone volume (TV) and bone contact with implant (TIC) is seen during the remodeling phase. In general, a transition period between the start of the remodeling (day 30) and the establishment of a steady state is observed. In the case of implant motion with 5 and 10 µm range, the bone-TV value is reduced from 100% and 95% to 60% and the transition period is predicted to be very long (over 40 years). On the other hand, in the case of 20 µm motion-range the bone-TV is reduced from approximately 60% to 20% within three years. Similar observations are made for the bone-TIC. However it should be noted that the bone-implant contact area (bone-TIC) is reduced faster than the total bone volume (bone-TV) during the remodeling phase, and larger drop in bone-TIC is seen as compared to bone-TV.

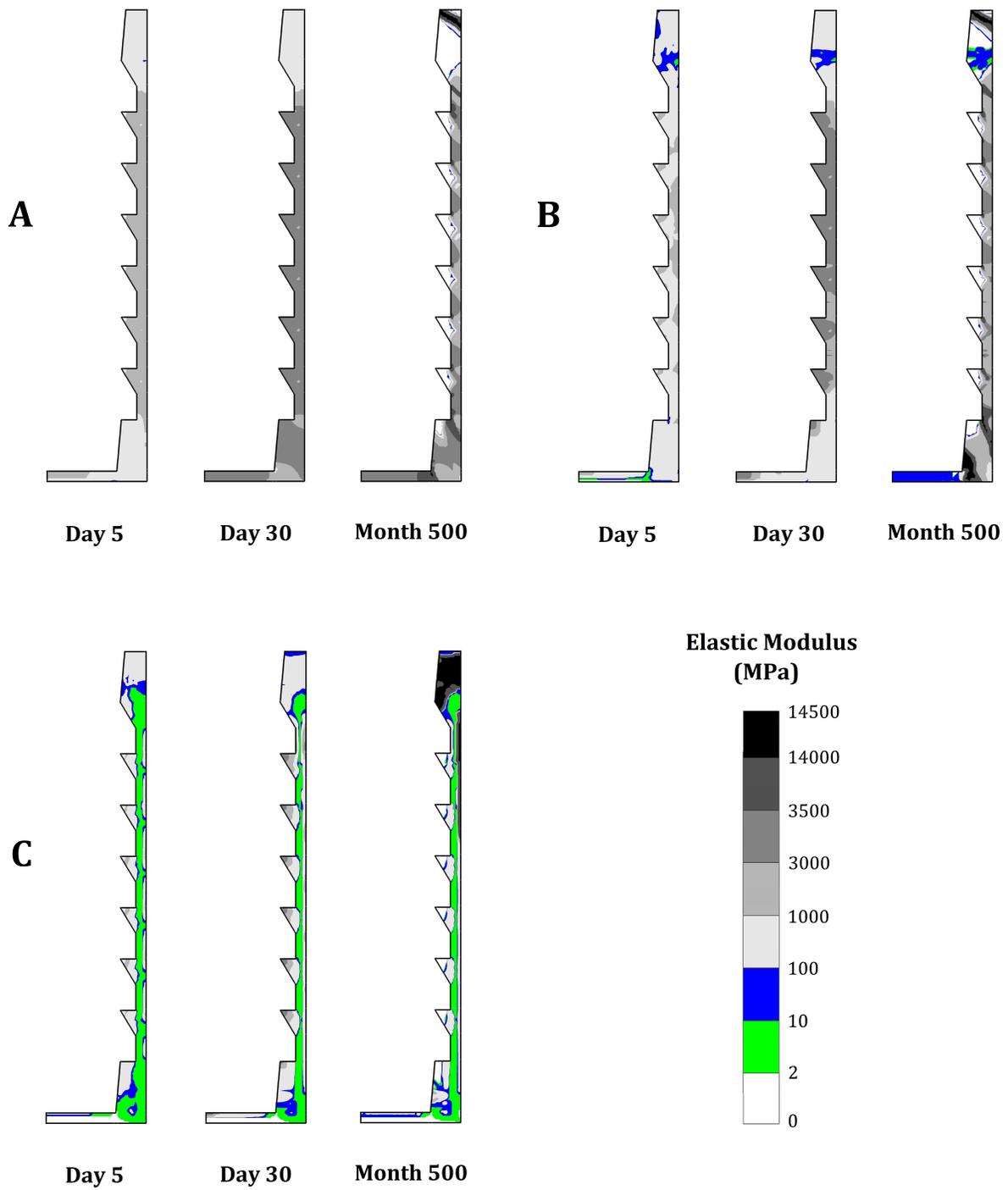

**Fig. 1** Transient change of local elastic modulus in the healing region for $z_{max}$ values of (**A**) 5 μm, (**B**) 10 μm and (**C**) 20 μm.

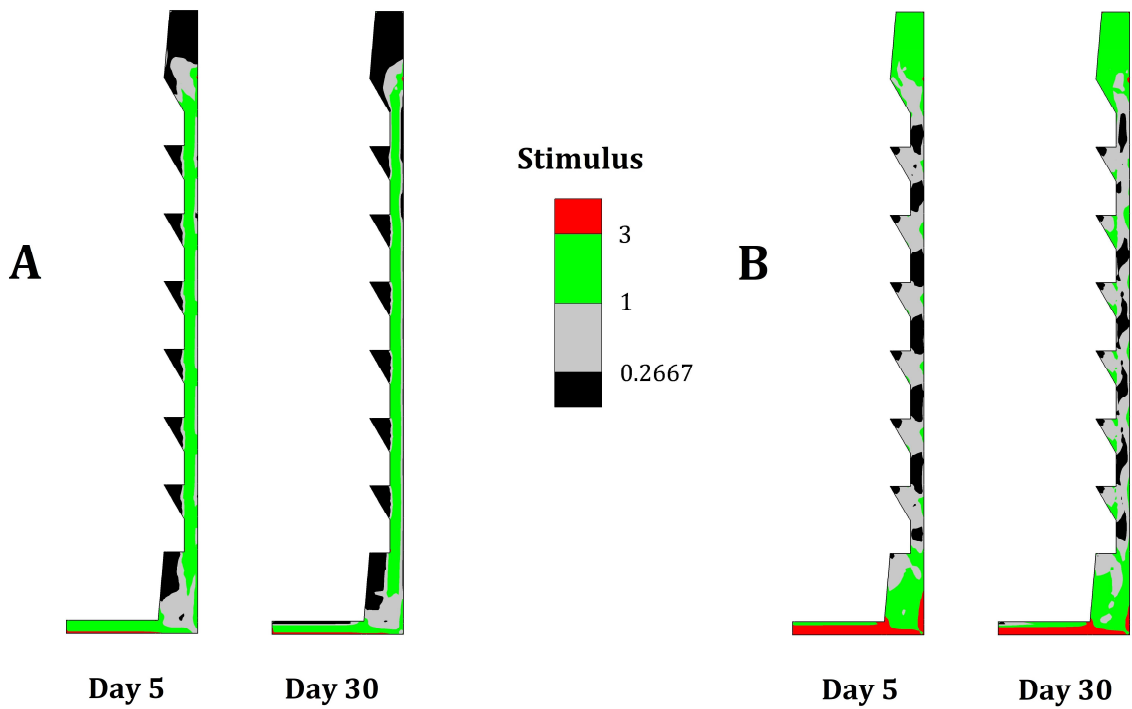

**Fig. 2** Distribution of (**A**) solid stimulus and (**B**) fluid stimulus in the healing region at day 5 and day 30 (healing phase) for a $z_{max}$ value of 20 μm.

The transient changes in the TV and TIC of cortical bone (E > 3.5 GPa) is shown in Fig. 5, only for the remodeling phase (day 30 and forward), as the healing phase does not result in bone that falls into this range of elastic modulus (Equation (3)). In fact, cortical bone develops from the immature and mature trabecular bone types that are presented in Fig. 4. It is noted that, regardless of micromotion amplitude, cortical bone volume (TV) is less than 15%. Interestingly, the trend is somewhat reversed in this plot, where the implant which was subjected to the highest micromotion (20 μm) during healing develops more cortical bone during remodeling. In particular, the bone –TV is on the order of 10 – 15% for the case of 20 μm, where as it is on the order of 2 – 7% for 5 and 10 μm. This is indicative of the bone remodeling process seeking a density distribution that can handle the applied load level.

Note that for the implant subjected to 20 μm, only 45% of the callus volume is available for remodeling by day-30 (Fig. 4). During remodeling 10 – 15% of the callus volume densifies to cortical bone (Fig. 5), whereas about 10% resorbs (Fig. 3). Cases with lower micromotion range (5 and 10 µm) result in less than 8% of cortical bone-TV. This is mostly due to more extensive resorption (Fig. 3) and partly due to existence of tissue with E < 100 MPa in 6% of the callus volume during the healing phase. Looking at the cortical bone-TIC, it is seen that all cases behave similarly and 8-10% of the implant contacts the cortical bone at the end of remodeling.

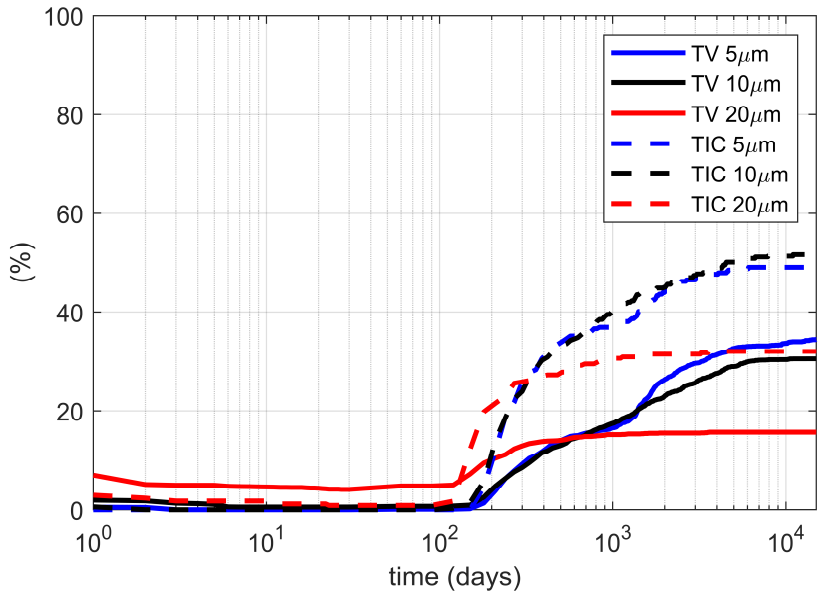

**Fig. 3** Percent of tissue in the healing volume (TV) and at the implant interface (TIC) with elastic modulus E < 2 MPa. This represents fibrous tissue during healing and resorbed tissue during bone during remodeling. Results are given for $z_{max}$ values of 5, 10 and 20 µm.

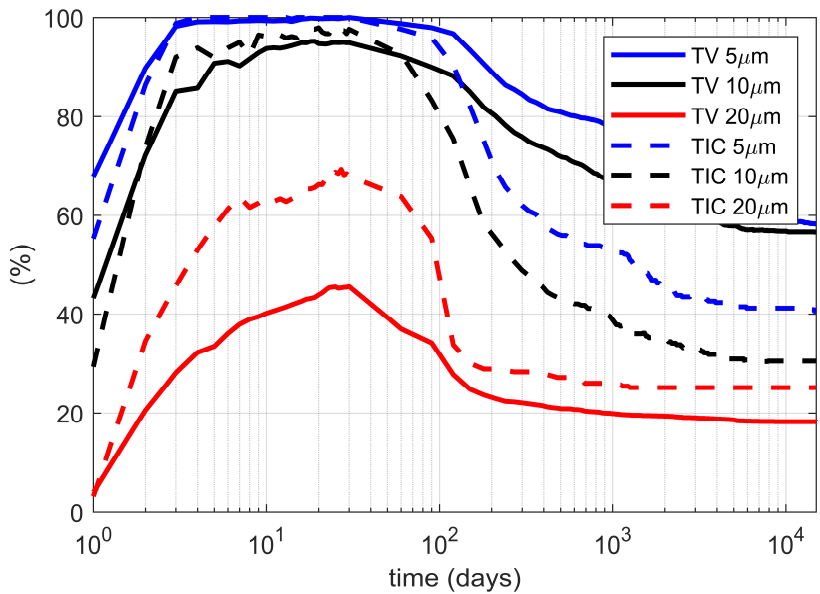

**Fig. 4** Percent of tissue in the healing volume (TV) and at the implant interface (TIC) with elastic modulus in the range of 100 MPa < E < 3.5 GPa. Results are given for $z_{max}$ values of 5, 10 and 20 μm.

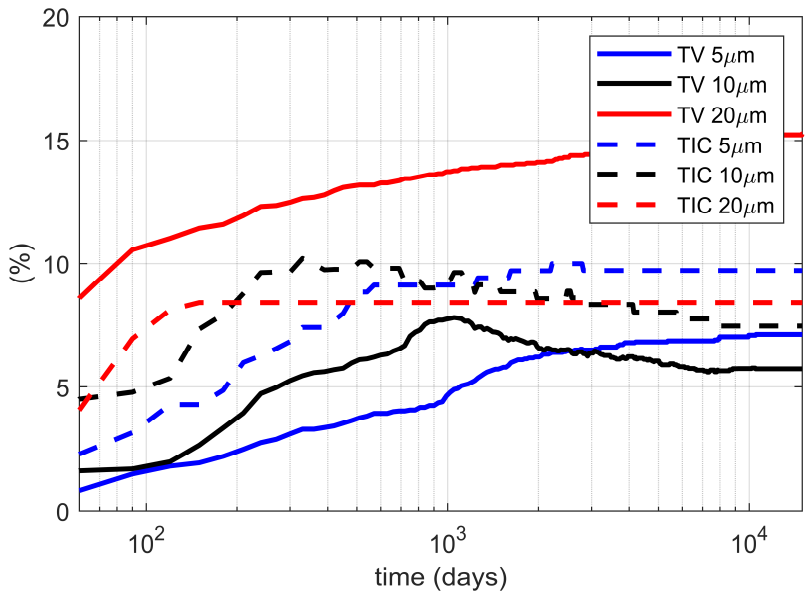

**Fig. 5** Percent of tissue in the healing volume (TV) and at the implant interface (TIC) with elastic modulus E > 3.5 GPa. Results are given for $z_{max}$ values of 5, 10 and 20 μm.

## DISCUSSION

In general, mechanical loading is in favor of formation of high-density bone during remodeling, but it is in favor of development of soft tissue during fracture healing. The fracture healing and bone remodeling theories both of which are rooted in empirical observation lead to this outcome. This work demonstrates this interplay between healing, remodeling and loading levels and shows that the point in time where bone quality is measured has a major role in the evaluation of the peri-implant osseointegration. This observation perhaps sheds light onto the seemingly contradictory results obtained in clinical and experimental studies.

There are numerous clinical and experimental studies showing the long-term success of implant treatment depends on the mechanical conditions applied to the implant during the healing phase. Sagara et al. attribute low levels of direct bone contact to early loading and excessive micromotion [29]. Piatelli et al. indicate immediate loading to be in favor of more bone implant contact (BIC) [30]. Henry et al. attribute more mature cortical bone around the implant to early loading [31].

Presence and quality of bone surrounding an implant [32] and its initial stability [3,4] have been extensively mentioned as important determinants of outcome of dental implant treatments. Excessive loading and relative motion of the implant are mentioned as important factors in development of interfacial fibrous tissue [32-34], which can be seen in Fig. 1C as well. To the contrary, Duyck et al. found that low micromotion is less favorable than a high micromotion [14]. This appears to be in agreement with evolution of TV in Fig. 5. Primary stability of the implant depends also on the insertion torque (IT) and the extent of initial BIC. High IT is expected to result in less implant micromotion. Cha et al. [16] showed that implants with high IT cause a wider zone of dying osteocytes at the implant interface, which is in agreement with the trend of TIC in Fig. 3. On the other hand, Grandi et al. showed that using high IT during the implant placement does not prevent osseointegration [15]. Simulation of the whole treatment process by using the tissue healing and bone remodeling processes sequentially, can explain the apparent inconsistencies reported in clinical studies. In particular, it is seen that reaching a bone mass distribution that appears favorable at the end of a three- or four-week long healing period may not be an indicator of the long-term bone maintenance. The entire healing and remodeling process should be considered to this end. On the other hand, as expected, this work confirms that a healing period that results in low quality/quantity is not indicative of long-term success/failure of the treatment.

**SUMMARY AND CONCLUSION**

Two dimensional, axisymmetric analysis of bone healing followed by bone remodeling was carried out in order to contribute to our understanding of long-term osseointegration and bone remodeling around early loaded dental implant systems. The work shows that evolution of tissue type following an implant treatment does not have a linear correlation with mechanical usage (i.e. micromotion levels). Moreover, the end state of tissue healing, which is the initial condition for bone remodeling, plays a crucial role in the final distribution of different tissue types around the implant. Without considering the tissue healing process, higher mechanical usage would guide the predictions toward a higher bone density and cannot predict development of soft tissue in the presence of excessive mechanical loading. On the other hand, studying only the tissue-healing phase does not provide any information about the long-term adaptation of internal bone density and potential regions of bone resorption. This work shows that an optimal range for implant micromotion and for a given implant contour should be possible, particularly on a patient specific basis, in order to achieve the desired outcome and functionality for dental implant treatments.

**MATERIALS AND METHODS**

In order to study peri-implant tissue healing and the subsequent bone remodeling, a 2D axisymmetric model (Fig. 6) of the bone and the implant was developed. The model includes a dental implant with inner and outer radius of 1.75 and 1.95 mm and height of 9 mm, cortical and cancellous bone regions, and the fracture callus. The healing region is 0.2 mm wide in which tissue properties evolve during healing and remodeling processes. In the healing phase, the bone and the tissue in the callus were modeled as poroelastic materials with the properties given in Table 1. Physics of a saturated porous medium is governed by fluid mass conservation as well as equations of elastic equilibrium [35]. In addition to the two material constants (elastic modulus and Poisson's ratio) defining the elastic behavior of the solid part of the healing region, four other material constants (dynamic permeability of the fluid, porosity, and solid and fluid bulk moduli) are required for the fluid mass conservation. The boundary conditions were defined as an ambient pore pressure (p=0) at the superior aspect of the cortical bone and the callus, and constrained displacements in the radial and axial directions representing the axisymmetric conditions (Fig. 6). Cyclical displacement-controlled loading was applied to the top of the implant along the implant (-z) axis. During the remodeling phase, all tissue types were assumed to behave in linear-elastic manner, where the physical deformation is governed solely by the equations of elastic equilibrium. The boundary conditions were kept the same as before. An axially oriented mastication load of 100 N was applied in -z direction.

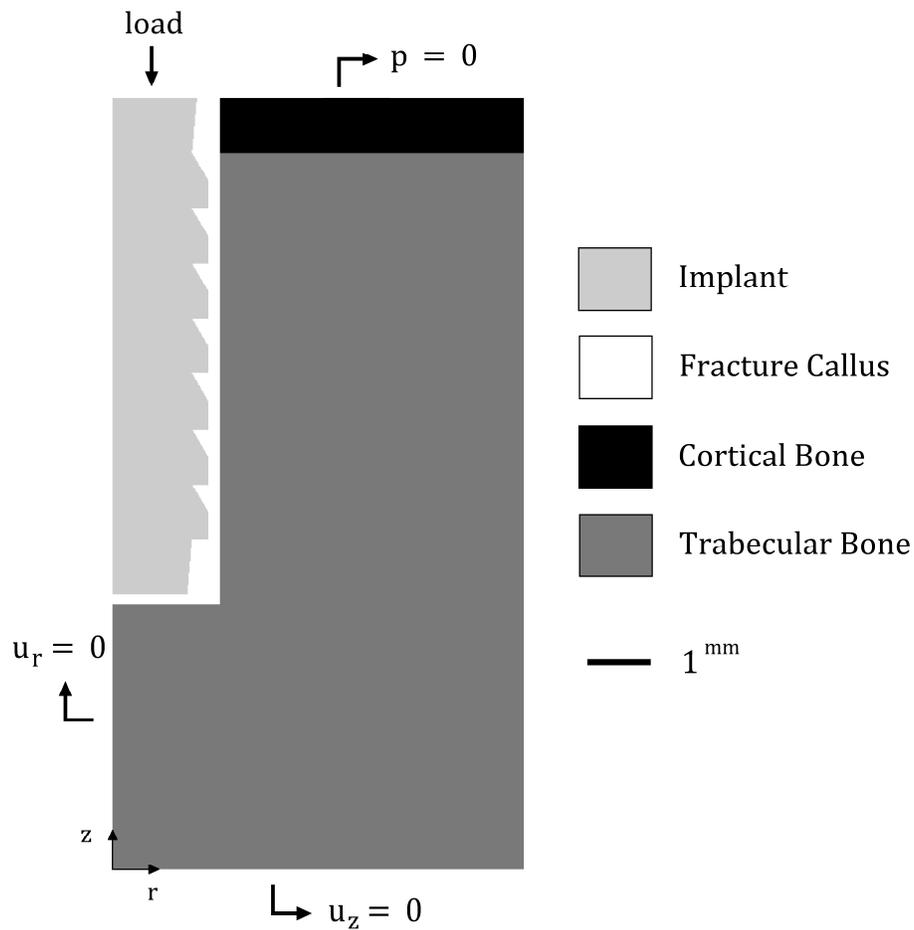

**Fig. 6** Cross-section depicting a dental implant with buttress type threads surrounded by cortical and cancellous bone types and the fracture callus.

**Fracture Healing**

Bone fracture healing is a physiologically complex process, which can go through various healing pathways based on the mechanical and biological factors [36]. The initial response to fracture in bone starts with migration of MSCs to the healing callus [37]. Lacroix and Prendergast suggested a random movement of stem cells (SCs) from a vascularized origin with maximum cell concentration toward the healing region [18], governed by the diffusion equation:

$$D\nabla^2 n_{sc} = \frac{dn_{sc}}{dt} \tag{1}$$

where D is the diffusion constant, $n_{sc}$ is local percentage of available stem cells, and t is the time. Note that at the beginning of fracture there are no stem cells in the healing region (i.e. $n_{sc} = 0$ at t = 0). D is calibrated such that $n_{sc}$ reaches its maximum ($n_{sc}^{(max)} = 100$) in the entire healing region after 14 days [38-40]. In this work $D = 0.023 \frac{mm^2}{day}$ was used.

Two biophysical stimuli, octahedral shear strain (γ) and interstitial fluid velocity (v) are thought to regulate cellular differentiation pathway [41,42]. The healing stimulus S is formulated as:

$$S = \frac{\gamma}{a} + \frac{v}{b} \tag{2}$$

where $a = 0.0375$ and $b = 3\,\mu m/s$ are two constants determined empirically [18]. Based on this regulatory model, higher values of S are described as the reason for fibrous tissue generation, while lower values of S predict bone tissue formation as follows:

| | | |
|---|---|---|
| Mature trabecular bone | $0.0000 < S < 0.2667$ | |
| Immature woven bone | $0.2667 < S < 1.0000$ | |
| Cartilaginous tissue | $1.0000 < S < 3.0000$ | (3) |
| Fibrous tissue | $3.0000 < S$ | |

The corresponding material properties of these tissue types are given in Table 1. During healing, the local cell concentration depends on diffusion. An effective value for a given material property $M_e$ in the healing region is found by using the rule of mixtures as follows:

$$M_e(t) = \frac{n_{sc}}{n_{sc}^{(max)}} M(t) + \left(1 - \frac{n_{sc}}{n_{sc}^{(max)}}\right) M_c \tag{4}$$

where $M_e$ is one of the six material properties listed in Table 1, $M(t)$ is the local poroelastic property that evolves as described by equation (3) and $M_c$ is the property for the callus tissue.

The evolution of material properties during healing depends on the concentration of available cells, the mechanical response of the poroelastic tissue and the stimulus described above. The time dependent nature of this coupled problem is solely due to the diffusion equation in this model. Equations (1) – (4) are solved through numerical iterations where each iteration is considered to be one day long. Due to the rapid changes in material properties during this pseudo-transient solution, numerical damping is introduced by using a moving average of the predicted quantities [18]. Recently we used the following relaxation approach as an alternative to the moving average in order to dampen the abrupt changes in material properties:

$$M_e^{(i+1)} = M_e^{(i)} + \alpha\left(M_e(t) - M_e^{(i)}\right), \qquad \alpha = 0.2 \tag{5}$$

where i represents the solution iteration level (i.e., day). Effects of initial material properties of the callus, geometrical properties, and MSC diffusion constant, on the healing pathway have been studied by Ghiasi et al. [43]. Another parametric study performed by the authors [25] shows that the D and α values used in the present work are effective in damping out the spurious fluctuations. In this work the material properties were updated in the fracture callus for 30 iterations, and final values were used as the initial conditions at the start of bone remodeling phase.

During fracture healing, the implant was subjected to oscillatory displacement according to the haversine function:

$$z(t) = \frac{1}{2} z_{max}(1 - \cos 2\pi\nu t) \tag{6}$$

where $z_{max}$ is the range of the micromotion, with an amplitude of $\pm z_{max}/2$. The oscillation frequency $\nu$ was kept constant at 1 Hz. $z_{max}$ values of 5, 10 and 20 μm were used in this work. Numerical experiments showed that simulation duration of 4 seconds was sufficient to find the steady state conditions. Average of fluid velocity and shear strain during transient solution was used to calculate healing stimulus in equation (2). The effects of loading rate, abrupt changes in loading conditions and calculation of the accumulated healing stimulus on the prediction of tissue healing pathway will be discussed in separate articles.

|  | Elastic Modulus (GPa) | Poisson's Ratio | Porosity | $K_s$ (GPa) | $K_f$ (GPa) | Permeability (mm$^4$/(N.s)) |
|---|---|---|---|---|---|---|
| Callus | 0.002 | 0.17 | 0.8 | 2.3 | 2.3 | 0.01 |
| Fibrous | 0.002 | 0.17 | 0.8 | 2.3 | 2.3 | 0.01 |
| Cartilage | 0.01 | 0.17 | 0.8 | 3.4 | 2.3 | 0.005 |
| Immature Woven Bone | 1 | 0.30 | 0.8 | 13.92 | 2.3 | 0.1 |
| Mature Trabecular Bone | 3.5 | 0.30 | 0.8 | 13.92 | 2.3 | 0.37 |
| Cortical Bone | 14.5 | 0.30 | 0.04 | 13.92 | 2.3 | $10^{-5}$ |

**Table 1** Poroelastic properties of tissues from Lacroix and Prendergast [18].

**Bone Remodeling**

Bone needs a certain level of mechanical stimulation to maintain a density distribution that can withstand the daily loading cycles [44]. The bone density does not change if the homeostatic stimulus level can be maintained. Loading levels that cause stimulus higher than the homeostatic level cause bone density to increase. At very high loading levels the bone can fracture. On the other hand, loading levels that result in stimulus lower than the homeostatic levels can cause the bone to resorb. These effects are carried out in two ways. Bone adapts both its shape (surface remodeling) and its internal material properties (internal remodeling) through cellular activities of osteoclasts removing dead bone cells and osteoblasts depositing new cells [45,46]. In the present study internal changes in bone material properties in response to the mechanical environment is investigated. Although it has been shown that bone remodeling depends on the fluid velocity as well as the mechanical signals in the solid phase [47], only the latter one was used in this work and bone and other tissue types were modeled as isotropic elastic materials. The effect of using a poroelastic model during bone remodeling should be considered in the future.

In nature and in this work, the end state of the healing phase serves as the initial condition of the remodeling phase. Assuming that soft tissue cannot remodel, only the healed regions with elastic modulus of 100 MPa and higher were allowed to remodel, or in other words experience density adaptation. This phase of adaptation was simulated by using Carter et al.'s model [44], which calculates the rate of change in bone density from the following relationship:

$$\dot{\rho} = \dot{r}\, S_v\, \rho^{(max)} \qquad (7)$$

where $\rho$ is bone density, $\rho^{(max)} = 1.92 \text{ gr/cm}^3$ is the maximum bone density and $S_v$ is bone specific surface (BS/TV) calculated from:

$$S_v = 0.6255\, \rho^6 - 3.703\, \rho^5 + 7.0228\, \rho^4 - 4.8345\, \rho^3 - 1.928\, \rho^2 + 6.745\, \rho \qquad (8)$$

In equation (7), $\dot{r}$ is the linear rate of bone apposition or resorption [44,48] that is represented as follows:

$$\dot{r} = \begin{cases} c_r(\psi - \psi_{AS}) + c_r\, w & \psi - \psi_{AS} < -w \\ 0 & -w < \psi - \psi_{AS} < w \\ c_f(\psi - \psi_{AS}) - c_f\, w & w < \psi - \psi_{AS} \end{cases} \qquad (9)$$

where $c_r = 2 \times 10^{-5}$ and $c_f = 2 \times 10^{-4}$ are the slope of the change in rate of bone resorption and bone apposition with respect to daily stimulus $\psi$. $\psi_{AS} = 15$ MPa/day is defined as attractor (or homeostatic) stress stimulus for 112 number of daily load cycles [44]. $w = 0.25\psi_{AS}$ is half width of the dead zone in which bone maintains its density ($\dot{r} = 0$). The daily stress stimulus is defined by using a tissue-level measure as follows:

$$\psi = (N\, \sigma_T^{\,m})^{1/m} \qquad (10)$$

where N is number of cycles of loading and in this work a value of $m = 4$ was used. The tissue level stress $\sigma_T$ is related to the continuum level stress $\sigma_c$ as follows:

$$\sigma_T = \left(\frac{\rho^{(max)}}{\rho}\right)^2 \sigma_c \qquad (11)$$

The continuum level stress can be represented by using the elastic modulus E and strain energy density u of the material as follows:

$$\sigma_c = \sqrt{2Eu} \tag{12}$$

Bone elastic modulus is related to its density [44] by the following empirical relationship:

$$E = \begin{cases} 2042.82\, \rho^{2.5}, & \rho < 1.2 \\ 1798.06\, \rho^{3.2}, & \rho > 1.2 \end{cases} \tag{13}$$

Equations (7)-(13) are solved numerically. In particular, equation (7) is discretized by using the forward time integration scheme. Each time step represents 30 days of bone loading and equation (10) is adjusted accordingly. The daily remodeling stimulus was determined by using N = 112 for a load value of 100 N on the tooth as the typical daily mastication regime.

**ACKNOWLEDGEMENTS**

This work was supported in part by Bicon Dental Implants (Boston, Massachusetts) through a research grant to Northeastern University.

**CONFLICT OF INTERESTS**

The authors do not have any conflict of interest.

**AUTHOR CONTRIBUTIONS**

Both authors conceived of the study and contributed to the writing. SI developed the computational algorithms and produced the results.

# REFERENCES

1. Jaffin, R. A., Kumar, A. & Berman, C. L. Immediate loading of implants in partially and fully edentulous jaws: a series of 27 case reports. Journal of periodontology **71**, 833-838, doi:10.1902/jop.2000.71.5.833 (2000).
2. Branemark, P.-I. Osseointegration and its experimental background. Journal of Prosthetic Dentistry **50**, 399-410, doi:10.1016/S0022-3913(83)80101-2 (1983).
3. Romanos, G. et al. Peri-implant bone reactions to immediately loaded implants. An experimental study in monkeys. Journal of periodontology **72**, 506-511, doi:10.1902/jop.2001.72.4.506 (2001).
4. Lioubavina-Hack, N., Lang, N. P. & Karring, T. Significance of primary stability for osseointegration of dental implants. Clinical oral implants research **17**, 244-250, doi:10.1111/j.1600-0501.2005.01201.x (2006).
5. Glauser, R. et al. Immediate Occlusal Loading of Brånemark Implants Applied in Various Jawbone Regions: A Prospective, 1-Year Clinical Study. Clinical implant dentistry and related research **3**, 204-213, doi:10.1111/j.1708-8208.2001.tb00142.x (2001).
6. Meyer, U. et al. Ultrastructural characterization of the implant/bone interface of immediately loaded dental implants. Biomaterials **25**, 1959-1967, doi:10.1016/j.biomaterials.2003.08.070 (2004).
7. Nkenke, E. et al. Bone contact, growth, and density around immediately loaded implants in the mandible of mini pigs. Clinical oral implants research **14**, 312-321, doi:10.1034/j.1600-0501.2003.120906.x (2003).
8. Brunski, J. B. In vivo bone response to biomechanical loading at the bone/dental-implant interface. Advances in dental research **13**, 99-119, doi:10.1177/08959374990130012301 (1999).
9. Prendergast, P. & Van der Meulen, M. in Bone Mechanics Handbook (ed Stephen C. Cowin) Ch. 32, 1-11 (2001).
10. Frost, H. Bone "mass" and the "mechanostat": a proposal. The Anatomical Record **219**, 1-9, doi:10.1002/ar.1092190104 (1987).
11. Hart, R. in Bone Mechanics Handbook (ed Stephen C. Cowin) Ch. 31, 1-42 (2001).
12. Wazen, R. M. et al. Micromotion-induced strain fields influence early stages of repair at bone-implant interfaces. Acta biomaterialia **9**, 6663-6674, doi:10.1016/j.actbio.2013.01.014 (2013).
13. Lin, D., Li, Q., Li, W. & Swain, M. Dental implant induced bone remodeling and associated algorithms. Journal of the mechanical behavior of biomedical materials **2**, 410-432, doi:10.1016/j.jmbbm.2008.11.007 (2009).
14. Duyck, J. et al. The influence of micro-motion on the tissue differentiation around immediately loaded cylindrical turned titanium implants. Archives of oral biology **51**, 1-9, doi:10.1016/j.archoralbio.2005.04.003 (2006).
15. Grandi, T., Guazzi, P., Samarani, R. & Grandi, G. Clinical outcome and bone healing of implants placed with high insertion torque: 12-month results from a multicenter controlled cohort study. International journal of oral and maxillofacial surgery **42**, 516-520, doi:10.1016/j.ijom.2012.10.013 (2013).
16. Cha, J. Y. et al. Multiscale analyses of the bone-implant interface. Journal of Dental Research **94**, 482-490, doi:10.1177/0022034514566029 (2015).
17. Vanegas-Acosta, J. & Garzón-Alvarado, D. A finite element method approach for the mechanobiological modeling of the osseointegration of a dental implant. Computer Methods and Programs in Biomedicine **101**, 297-314, doi:10.1016/j.cmpb.2010.11.007 (2011).

18	Lacroix, D. & Prendergast, P. A mechano-regulation model for tissue differentiation during fracture healing: analysis of gap size and loading. Journal of biomechanics **35**, 1163-1171, doi:10.1016/S0021-9290(02)00086-6 (2002).
19	Crupi, V., Guglielmino, E., La Rosa, G., Vander Sloten, J. & Van Oosterwyck, H. Numerical analysis of bone adaptation around an oral implant due to overload stress. Proceedings of the Institution of Mechanical Engineers, Part H: Journal of Engineering in Medicine **218**, 407-415, doi:10.1243/0954411042632171 (2004).
20	Shefelbine, S. J., Augat, P., Claes, L. & Simon, U. Trabecular bone fracture healing simulation with finite element analysis and fuzzy logic. Journal of biomechanics **38**, 2440-2450, doi:10.1016/j.jbiomech.2004.10.019 (2005).
21	Li, J. et al. A mathematical model for simulating the bone remodeling process under mechanical stimulus. Dental Materials **23**, 1073-1078, doi:10.1016/j.dental.2006.10.004 (2007).
22	Chou, H. Y., Jagodnik, J. J. & Muftu, S. Predictions of bone remodeling around dental implant systems. Journal of biomechanics **41**, 1365-1373, doi:10.1016/j.jbiomech.2008.01.032 (2008).
23	Chou, H.-Y., Romanos, G., Müftü, A. & Müftü, S. Peri-implant bone remodeling around an extraction socket: predictions of bone maintenance by finite element method. The International Journal of Oral & Maxillofacial Implants **27**, e39-48 (2012).
24	Chou, H. Y. & Muftu, S. Simulation of peri-implant bone healing due to immediate loading in dental implant treatments. Journal of biomechanics **46**, 871-878, doi:10.1016/j.jbiomech.2012.12.023 (2013).
25	Irandoust, S. & Muftu, S. in Northeast Bioengineering Conference (NEBEC), 2014 40th Annual.  1-2 (IEEE, 2014).
26	Irandoust, S. & Muftu, S. in Biomedical Engineering Conference (NEBEC), 2015 41st Annual Northeast.  1-2 (IEEE).
27	Chou, H.-Y. & Müftü, S. Corrigendum to "Simulation of peri-implant bone healing due to immediate loading in dental implant treatments"[J. Biomech. 46/5 (2013) 871–878]. Journal of biomechanics **49**, 2000-2005, doi:10.1016/j.jbiomech.2016.04.014 (2016).
28	Limbert, G. et al. Trabecular bone strains around a dental implant and associated micromotions--a micro-CT-based three-dimensional finite element study. Journal of biomechanics **43**, 1251-1261, doi:10.1016/j.jbiomech.2010.01.003 (2010).
29	Sagara, M., Akagawa, Y., Nikai, H. & Tsuru, H. The effects of early occlusal loading on one-stage titanium alloy implants in beagle dogs: a pilot study. The Journal of Prosthetic Dentistry **69**, 281-288, doi:10.1016/0022-3913(93)90107-Y (1993).
30	Piattelli, A., Corigliano, M., Scarano, A., Costigliola, G. & Paolantonio, M. Immediate loading of titanium plasma-sprayed implants: an histologic analysis in monkeys. Journal of periodontology **69**, 321-327, doi:10.1902/jop.1998.69.3.321 (1998).
31	Henry, P. J., Tan, A., Leavy, J., Johansson, C. B. & Albrektsson, T. Tissue regeneration in bony defects adjacent to immediately loaded titanium implants placed into extraction sockets: a study in dogs. The International Journal of Oral & Maxillofacial Implants **12**, 758-766 (1996).
32	Meyer, U., Vollmer, D., Runte, C., Bourauel, C. & Joos, U. Bone loading pattern around implants in average and atrophic edentulous maxillae: a finite-element analysis. Journal of Cranio-Maxillofacial Surgery **29**, 100-105, doi:10.1054/jcms.2001.0198 (2001).
33	Brunski, J. B. The influence of force, motion and related quantities on the response of bone to implants. Non-Cemented Total Hip Arthroplasty **43** (1988).
34	Søballe, K., Hansen, E. S., B-Rasmussen, H., Jørgensen, P. H. & Bünger, C. Tissue ingrowth into titanium and hydroxyapatite-coated implants during stable and unstable mechanical conditions. Journal of Orthopaedic Research **10**, 285-299, doi:10.1002/jor.1100100216 (1992).

35    Wang, H. F. Theory of linear poroelasticity with applications to geomechanics and hydrogeology. (Princeton University Press, 2017).
36    Ghiasi, M. S., Chen, J., Vaziri, A., Rodriguez, E. K. & Nazarian, A. Bone fracture healing in mechanobiological modeling: A review of principles and methods. Bone Reports **6**, 87-100, doi:10.1016/j.bonr.2017.03.002 (2017).
37    Shabbir, A., Cox, A., Rodriguez-Menocal, L., Salgado, M. & Badiavas, E. V. Mesenchymal Stem Cell Exosomes Induce Proliferation and Migration of Normal and Chronic Wound Fibroblasts, and Enhance Angiogenesis In Vitro. Stem Cells and Development, doi:10.1089/scd.2014.0316 (2015).
38    Einhorn, T. A. The cell and molecular biology of fracture healing. Clinical Orthopaedics and Related Research **355**, S7-S21 (1998).
39    Dimitriou, R., Tsiridis, E. & Giannoudis, P. V. Current concepts of molecular aspects of bone healing. Injury **36**, 1392-1404, doi:10.1016/j.injury.2005.07.019 (2005).
40    Nakamizo, A. et al. Human bone marrow–derived mesenchymal stem cells in the treatment of gliomas. Cancer Research **65**, 3307-3318, doi:10.1158/0008-5472.CAN-04-1874 (2005).
41    Huiskes, R., Van Driel, W., Prendergast, P. & Søballe, K. A biomechanical regulatory model for periprosthetic fibrous-tissue differentiation. Journal of Materials Science: Materials in Medicine **8**, 785-788, doi:10.1023/A:1018520914512 (1997).
42    Prendergast, P., Huiskes, R. & Søballe, K. Biophysical stimuli on cells during tissue differentiation at implant interfaces. Journal of biomechanics **30**, 539-548, doi:10.1016/S0021-9290(96)00140-6 (1997).
43    Ghiasi, M. S., Chen, J. E., Rodriguez, E. K., Vaziri, A. & Nazarian, A. Computational Modeling of the Effects of Inflammatory Response and Granulation Tissue Properties on Human Bone Fracture Healing. arXiv preprint arXiv:1808.04458 (2018).
44    Carter, D. R. & Beaupré, G. S. in Skeletal function and form: mechanobiology of skeletal development, aging, and regeneration    138-160 (Cambridge University Press, 2007).
45    Cowin, S. C. & Van Buskirk, W. Surface bone remodeling induced by a medullary pin. Journal of biomechanics **12**, 269-276, doi:10.1016/0021-9290(79)90069-1 (1979).
46    Cowin, S. & Van Buskirk, W. Internal bone remodeling induced by a medullary pin. Journal of biomechanics **11**, 269-275, doi:10.1016/0021-9290(78)90053-2 (1978).
47    Pereira, A. F., Javaheri, B., Pitsillides, A. & Shefelbine, S. Predicting cortical bone adaptation to axial loading in the mouse tibia. Journal of the Royal Society Interface **12**, 20150590, doi:10.1098/rsif.2015.0590 (2015).
48    Beaupré, G., Orr, T. & Carter, D. An approach for time-dependent bone modeling and remodeling—theoretical development. Journal of Orthopaedic Research **8**, 651-661, doi:10.1002/jor.1100080506 (1990).